\documentclass[lettersize, journal]{IEEEtran}
\IEEEoverridecommandlockouts
\usepackage{caption}
\usepackage{subcaption}
\usepackage{cite}
\usepackage{amsmath,amssymb,amsfonts}
\usepackage[linesnumbered,ruled,vlined]{algorithm2e}
\usepackage{multicol}
\usepackage{algpseudocode}  
\usepackage{graphicx} 
\usepackage{textcomp}
\usepackage{xcolor}
\usepackage{adjustbox}
\usepackage{chngcntr}
\usepackage[hyphens]{url}
\usepackage{csquotes}
\usepackage{hyperref}
\usepackage{multirow}
\hypersetup{breaklinks=true}
\counterwithout{table}{section}
\ifCLASSINFOpdf
\else
\fi
\begin{document}
\markboth{IEEE Transactions on Service Computing}%
{Shell \MakeLowercase{\textit{et al.}}: Bare Demo of IEEEtran.cls for IEEE Journals}
\title{DRPC: Distributed Reinforcement Learning Approach for Scalable Resource Provisioning in Container-based Clusters

\author{Haoyu Bai
\IEEEmembership{Fellow, IEEE}, Masaki Owari
\thanks{M. Hayashi is with Graduate School 
of Mathematics, Nagoya University, Nagoya, 
Japan}
\thanks{M. Owari is with the Faculty of 
Informatics, Shizuoka University, 
Hamamatsu, Shizuoka, Japan.}
}

\author{Haoyu Bai,
        Minxian Xu,~\IEEEmembership{Senior Member,~IEEE,}
        Kejiang Ye,~\IEEEmembership{Senior Member,~IEEE,}
        Rajkumar Buyya,~\IEEEmembership{Fellow,~IEEE,}
        Chengzhong Xu,~\IEEEmembership{Fellow,~IEEE}%

\thanks{H. Bai and R. Buyya are with the School of Computing and Information Systems, the University of Melbourne, Melbourne, Australia. H. Bai contributed this work when he was a visiting student at Shenzhen Institute of Advanced Technology, Chinese Academy of Sciences. 

M. Xu and K. Ye are with Shenzhen Institute of Advanced Technology, Chinese Academy of Sciences, Shenzhen, China. 

C. Xu is with State Key Lab of IOTSC, University of Macau, Macau, China. 


This work is supported by the National Key R\&D Program of
China (No.2021YFB3300200), the National Natural Science Foundation
of China (No. 62072451, 62102408), Shenzhen Industrial Application
Projects of undertaking the National key R \& D Program of China (No.
CJGJZD20210408091600002), and Shenzhen Science and Technology Program
(No. RCBS20210609104609044).

M. Xu is the corresponding author.}
}

}

\maketitle

\begin{abstract}



Microservices have transformed monolithic applications into lightweight, self-contained, and isolated application components, establishing themselves as a dominant paradigm for application development and deployment in public clouds such as Google and Alibaba. Autoscaling emerges as an efficient strategy for managing resources allocated to microservices' replicas. However, the dynamic and intricate dependencies within microservice chains present challenges to the effective management of scaled microservices. Additionally, the centralized autoscaling approach can encounter scalability issues, especially in the management of large-scale microservice-based clusters. To address these challenges and enhance scalability, we propose an innovative distributed resource provisioning approach for microservices based on the Twin Delayed Deep Deterministic Policy Gradient algorithm. This approach enables effective autoscaling decisions and decentralizes responsibilities from a central node to distributed nodes. Comparative results with state-of-the-art approaches, obtained from a realistic testbed and traces, indicate that our approach reduces the average response time by 15\% and the number of failed requests by 24\%, validating improved scalability as the number of requests increases.
\end{abstract}

\begin{IEEEkeywords}
Cloud computing, Kubernetes, microservice, reinforcement learning, distributed resources management
\end{IEEEkeywords}

\section{introduction}
Microservice architecture has emerged as a transformative approach in the field of software design and development. It is characterized by its modular and decentralized structure, where complex applications are broken down into smaller, independently deployable services\cite{Newman_2021}\cite{GiovanniTSC2024}. Each microservice focuses on a specific business capability, allowing teams to work on distinct components simultaneously and enabling rapid development, deployment, and scalability. This architecture promotes flexibility, resilience, and maintainability by minimizing the impact of changes within one service on the overall system \cite{Microsoft2021}. As microservices communicate through well-defined APIs, they facilitate seamless integration and support heterogeneous technology stacks. 
The cloud service providers, such as Amazon, Google, Microsoft, and Alibaba, have embraced microservice to develop and deploy their applications in cloud computing environment \cite{SPE2024}.

The rapid growth of the microservices paradigm has introduced challenges in resource management \cite{pallewatta2023placement}. As the number of microservices increases within a system, ensuring optimal resource allocation and utilization becomes complex. Microservices autoscaling has emerged as a viable solution to address these challenges \cite{jawaddi2022review}. Autoscaling involves dynamically adjusting the number of instances or resources allocated to microservices based on real-time demand, ensuring efficient resource utilization while maintaining desired performance levels \cite{XieTSC2024}. This technique encompasses both horizontal scaling, which involves adding or removing replicas of a service, and vertical scaling, which involves adjusting the resources allocated to each instance. Autoscaling mechanisms utilize various metrics and triggers, such as CPU usage, memory consumption, and request rates, to make decisions about scaling actions \cite{kalavri2018three}. Implementing effective autoscaling strategies enhances system responsiveness, minimizes operational costs, and optimizes resource allocation in dynamic and unpredictable environments.

Moreover, the intricate interdependencies among microservices often give rise to critical paths and key nodes that significantly impact performance. Efficiently implementing autoscaling for microservices is confronted with its own set of challenges and intricacies. The dynamic nature of these dependencies necessitates a profound understanding of the application's behavior, workload patterns, and their implications for system performance. 
Identifying the optimal scaling strategy, striking a balance in resource allocation across interconnected services while maintaining overall system stability, proves to be a non-trivial undertaking \cite{luo2023optimizing}. Additionally, the real-time nature of scaling decisions and the imperative to minimize disruptions to service quality and user experience further contribute to the complexity.

Reinforcement learning (RL) \cite{padakandla2021survey} has the potential to optimize the scaling of microservices, \color{black}taking into account their intricate interdependencies. Nevertheless, the majority of RL solutions adopt a centralized decision-making approach, which poses scalability challenges. This centralized structure encounters difficulties with the increasing complexity and interdependencies of microservices, resulting in adverse effects on scalability, response times, and reliability. 
Centralized management approaches such as Borg \cite{verma2015large} permit users to over-provision resources while assigning jobs to machines, leading to resource wastage and consequently diminishing overall performance in resource-limited environments. In addition, centralized approaches can compromise system performance as microservices scale up, given the continuous communication between huge amount of services and the central node may introduce bottlenecks and latency. 

In this paper, we explore novel approaches to address these limitations and propose a distributed reinforcement learning framework for microservices automatic scaling. This framework aims to enhance scalability, improve decision-making efficiency, and ensure the adaptability of microservices scaling strategies to dynamic and evolving environments. By decentralizing the decision-making process, we aim to mitigate the challenges associated with centralized approaches and unlock the full potential of RL in optimizing microservices resource allocation and scalability.

The main \textbf{contributions} of this paper can be summarized as follows:

\begin{itemize}

\item We design a \underline{d}istributed \underline{r}einforcement learning framework for resource \underline{p}rovisioning for \underline{c}ontainer-based autoscaling (\textit{DRPC}). This framework facilitates precise modelling of system resources and their scalable allocation to address the dynamic demands of microservices.

\item We propose a method for selecting optimization strategies based on reinforcement learning and a distributed algorithm that incorporates domain knowledge through deep imitation learning. This approach facilitates efficient and adaptive decision-making, optimizing the choice of scaling strategies across clusters of microservices.

\item  We perform comprehensive testing and validation of our proposed model and policies, utilizing real-world traces and a dedicated testing platform. These experiments are conducted to thoroughly assess the effectiveness and performance of our distributed reinforcement learning framework for microservices autoscaling.
 \end{itemize}

 
\section{Related Work}
In this section, we discuss the current autoscaling methods for microservices, categorizing them into three groups according to their key mechanisms: threshold-based and heuristic approaches, machine learning (ML) based approaches, and deep learning (DL) based approaches.
 
\label{related_work}

\subsection{Threshold-based and Heuristic Autoscaling}

The threshold-based heuristic approach for microservice resource allocation relies on predefined rules, scaling resources up when utilization surpasses a predetermined threshold (e.g., 75\%). This method proves efficient in scenarios with abundant resources and stable request patterns.  
 He et al. \cite{he2019re} leveraged both genetic and heuristic algorithms to determine the optimized microservice deployment location within an edge-cloud environment.  Horizontal pod auto-scaler (HPA) \cite{burns2019kubernetes} is a scaling technique adopted in Kubernetes, primarily focusing on horizontal scaling. It is capable of dynamically adjusting the number of replicas based on resource variations, such as CPU and memory, within the existing servers. The primary objective of HPA is to determine the number of replicas to be added or removed based on system running status. Kannan et al. \cite{Kannan2019} conceptualized multi-stage tasks using a Directed Acyclic Graph (DAG). By leveraging the DAG, they could predict the task's completion duration. Their method continually adapts thresholds and algorithms for each microservice. As a result, every microservice can manage its loads autonomously, incurring minimal communication expenses.

\subsection{Machine Learning based Autoscaling}
The ML-based autoscaling for microservices dynamically adjusts resources through ML algorithms that analyze historical data and workload patterns. This method optimizes resource utilization and ensures consistent performance by scaling services as required.  
Hou et al. \cite{hou2020ant} proposed an autoscaling approach that emphasizes both power and latency considerations for resource provisioning at micro and macro levels. By employing decision trees and a tagging methodology, they were able to expedite the resource-matching process. Yu et al. \cite{Yu2019} introduced a framework called Microscaler that utilizes a service mesh to monitor resource usage patterns. By integrating online learning and heuristic methods, the framework achieves near-optimal solutions to satisfy resource needs and quality of service (QoS) requirements. Liu et al. \cite{liu2019qos} undertook a detailed analysis of bottlenecks in prevalent microservice applications and incorporated various machine learning models to facilitate resource scheduling. Meanwhile, Gan et al. \cite{gan2019seer} harnessed predictive methodologies to detect QoS violations. They leveraged a combination of machine learning models and expansive historical data to pinpoint the microservices responsible for these QoS infractions, enabling better resource reallocation to mitigate QoS degradation.

\begin{table*}[ht]
		\centering
		\caption{Comparison of related work}
		\label{tab:relatedwork}
		\resizebox{1.0\textwidth}{!}{
			\begin{tabular}{|c|c|c|c|c|c|c|c|c|c|c|}
				\hline
				\multirow{2}*{\textbf{Approach}} & \multicolumn{3}{c|}{\textbf{Scaling Techniques}} & \multicolumn{2}{c|}{\textbf{Workload Prediction}} & \multicolumn{3}{c|}{\textbf{Scheduling Mechanism}} & \multicolumn{2}{c|}{\textbf{Decision-making Pattern}}\\ \cline{2-11} 
				& \textbf{Vertical} & \textbf{Horizontal} & \textbf{Brownout} & \textbf{Linear} & \textbf{Non-Linear} & \textbf{Heuristic} & \textbf{Machine Learning}  & \textbf{Deep Learning}  & \textbf{Distributed} & \textbf{Centralised}  \\ \hline
                He et al.  \cite{he2019re}              & $\checkmark$ & $\checkmark$ &  &         &         & $\checkmark$ &        &        &$\checkmark$ &        \\ \hline
				Burns et al. \cite{burns2019kubernetes} &         & $\checkmark$ &         &         &         & $\checkmark$ &        &        &        &$\checkmark$ \\ \hline
				Kannan et al. \cite{Kannan2019}         & $\checkmark$ &         &         & $\checkmark$ &         & $\checkmark$ &        &        &        &$\checkmark$ \\ \hline
				Hou et al. \cite{hou2020ant}            & $\checkmark$ &         &         & $\checkmark$ &         &         &$\checkmark$ &        &        & $\checkmark$ \\ \hline
				Yu et al. \cite{Yu2019}                 &         & $\checkmark$ &         &         & $\checkmark$ &         &$\checkmark$ &        &        & $\checkmark$ \\ \hline
				Liu et al. \cite{liu2019qos}            & $\checkmark$ &         &         &         & $\checkmark$ &         & $\checkmark$&        &        &$\checkmark$ \\ \hline
				Gan et al. \cite{gan2019seer}           & $\checkmark$ &         &         &         & $\checkmark$ &         &$\checkmark$ &        &        &$\checkmark$ \\ \hline
				Rzadca et al. \cite{rzadca2020autopilot}& $\checkmark$ & $\checkmark$ &         & $\checkmark$ &         &         &        &$\checkmark$ &        &$\checkmark$  \\ \hline
				Xu et al. \cite{xu2022CoScal}           & $\checkmark$ & $\checkmark$ & $\checkmark$ &         & $\checkmark$ &         &        &$\checkmark$ &        &$\checkmark$  \\ \hline
				Qiu et al. \cite{qiu2020firm}           & $\checkmark$ &         &         &         &         &         &        &$\checkmark$ &        &$\checkmark$ \\ \hline
				Zhang et al. \cite{Zhang2020ASARSA}     &         & $\checkmark$ &         &         &         &         &        &$\checkmark$ &        &$\checkmark$ \\ \hline
				Rossi et al. \cite{Rossi2019}           & $\checkmark$ & $\checkmark$ &         & $\checkmark$ &         &         &        &$\checkmark$ &        &$\checkmark$ \\ \hline
                Our method (DRPC)                             & $\checkmark$ & $\checkmark$ & $\checkmark$ &         & $\checkmark$ &         &        &$\checkmark$ &$\checkmark$ &        \\  \hline    
			\end{tabular} 
		}
	\end{table*}

\subsection{Deep Learning based Autoscaling}

The emphasis has shifted towards employing DL for microservice autoscaling, utilizing neural networks for decision-making and pattern analysis. This approach dynamically adjusts resource allocation based on real-time demand and optimizes provisioning by learning from historical and current data, thereby minimizing resource wastage.
Autopilot \cite{rzadca2020autopilot} accumulates historical server data and then utilizes two scheduling techniques: the sliding window algorithm based on past data, and the meta-algorithm inspired by RL. CoScal \cite{xu2022CoScal} classifies resource usage into four representative levels. It then applies an approximated Q-learning algorithm to these segments, establishing an estimated policy. \color{black}This policy encompasses horizontal and vertical scaling, as well as  brownout \cite{XuCSUR2019} capabilities that can dynamically activate and deactivate application component.  With gated recurrent unit (GRU), CoScal predicts upcoming workloads and refers to the trained lookup table to determine the suitable scaling action for each pod, ensuring optimization across all pods. \color{black} FIRM \cite{qiu2020firm} employs a systematic approach to allocating resources in cloud systems. It pinpoints the crucial path in the microservice dependency by carefully examining the connections between components. Once this path is identified, FIRM employs a specialized Support Vector Machine (SVM) tailored to operate on both a per-critical-path and per-micro-service-instance basis. This helps in identifying specific microservice instances that require optimization. Zhang et al. \cite{Zhang2020ASARSA} introduced a predictive RL algorithm for horizontal container scaling, which combines the Autoregressive Integrated Moving Average (ARIMA) model with a neural network model, ensures the predictability and precision of the scaling procedure. Rossi et al. \cite{Rossi2019} developed RL-based strategies to manage both horizontal and vertical scaling for containers. This approach enhances system adaptability in the face of fluctuating workloads and hastens the learning phase by harnessing varying levels of environmental knowledge. 

\subsection{Critical Analysis}

Although the existing representative methods have brought valuable contributions, our proposed method advances the relevant area in several key points. First, unlike HPA and other threshold-based approaches that primarily focus on horizontal scaling, our approach leverages multi-faceted scaling approaches (horizontal, vertical and brownout that can dynamically activate or deactivate optional microservices \cite{XuCSUR2019}), ensuring optimal resource allocation even during non-overload-states. Second, compared with the machine learning-based approach, we have applied deep learning to capture the features of workloads and utilized RL to make scaling decisions to achieve more accurate and efficient resource provisioning decisions. The compared differences with the related work are highlighted in Table \ref{tab:relatedwork}.

Our approach is most similar to CoScal \cite{xu2022CoScal} and FIRM \cite{qiu2020firm} based on RL to auto-scale microservices while having significant differences compared with them. 
In contrast to CoScal's large-grained segmentation of system states, our method employs a deep neural network, offering nuanced decision-making and the capability for simultaneous multiple scaling actions, crucial for sudden load changes. Compared to FIRM, which focuses on optimizing resources of the critical path and nodes, our method ensures comprehensive optimization, with the advantage of supporting both horizontal and vertical scaling, and brownout. Furthermore, our advanced resource allocation framework provides precise solutions, overcoming the scalability limitations inherent in FIRM's centralized reinforcement learning approach, especially under a high volume of requests.


\section{Motivation}
\label{sec:motivation}

\color{black}In this section, we perform motivational experiments with a use case that requests increase quickly to explore the system scalability under a centralized design, such as the centralized database capturing the dynamic change of microservices (FIRM) \cite{qiu2020firm} and the centralized RL-based approach (CoScal) \cite{xu2022CoScal}. \color{black}The investigation focuses on response time as the number of requests significantly increases. For these experiments, we utilize the TrainTicket \cite{TrainTicket} microservice application comprising approximately 40 services, including user, station, price, and route. The experiments are conducted on two nodes: one for the initial TrainTicket deployment and another for scaling, with a configuration of up to 8 CPU cores and 8 GB memory.  

\begin{figure}[htbp]
\centerline{\includegraphics[width=0.8\linewidth]{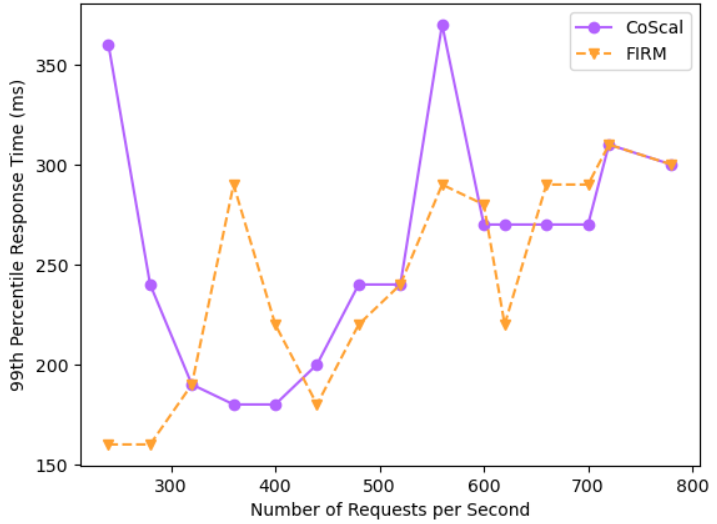}}
\caption{Response time of centralized approaches when the number of requests increases significantly}
\label{fig:ResponsetimeVSRequestRate}
\end{figure}

\color{black}Fig. \ref{fig:ResponsetimeVSRequestRate} shows the performance of FIRM and CoScal when the number of requests increases from 200 to 800 per second within a short time (e.g. 1 minute), and we can notice apparent performance fluctuations. \color{black}For instance, for the FIRM approach, the response time increases from 160 ms to 290 ms when requests increase from 200 to 400 per second. This resulted from the limitation of the centralized design of FIRM in that its central node contains functionalities including data collection, training, inference, and service provisioning. The CoScal based on centralized RL design also suffers from the same issue when the number of requests increases significantly within a short time, CoScal might over-provision into the system to ensure response time but also incurs resource wastage. For instance, several replicas are added to provision resources when the number of requests increases and the response time is also reduced. 

\color{black}
In this work, our objective is to provision resources efficiently via an RL-based approach, and utilize a distributed framework to overcome the limitation of centralized design, e.g. system performance bottleneck. A centralized module responsible for data collection and inference can exhibit drawbacks such as unreliability under a container-based environment, slow performance, and inability to adjust system resources asynchronously. The system could also potentially fail to adjust resources if the central nodes experience performance degradation. Hence, we propose a distributed framework utilizing multiple lightweight neural networks on distributed nodes to execute the operations sent from the central node to relieve the burden of the central node. This approach can potentially hasten resource allocation, and provide a more accurate resource usage prediction of the cloud system's behaviour. Moreover, it could facilitate differential resource adjustment, accommodating services with rapid changes more frequently than others. In the subsequent sections, we will introduce our detailed solution. 
\color{black}

\section{System model}
\label{system_model}

\begin{figure*}[htbp]
\centerline{\includegraphics[width=0.75\linewidth]{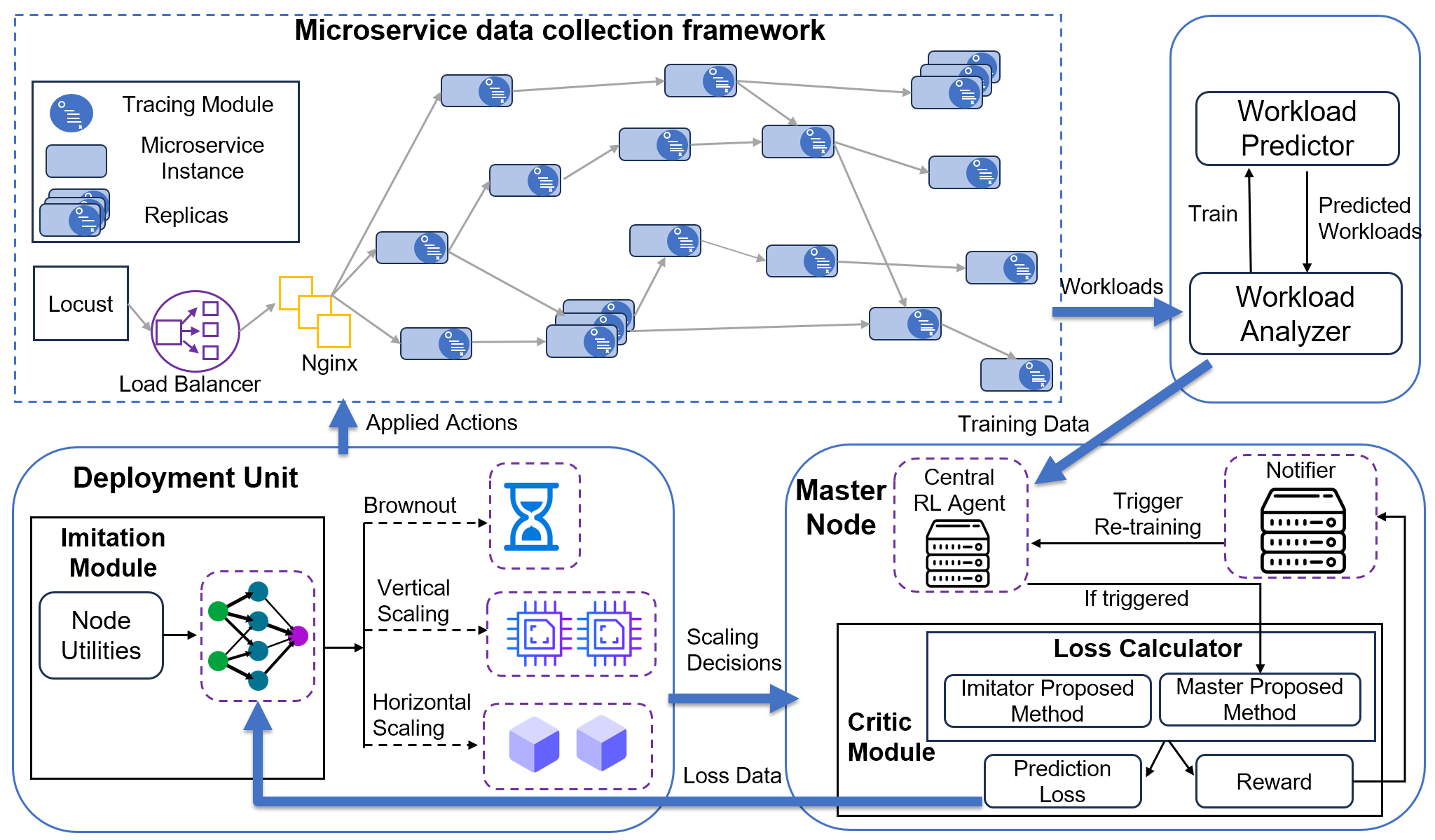}}
\caption{System Model of DRPC}
\label{fig:SystemModel}
\end{figure*}

In this section, we will introduce the system model of DRPC, which adheres to the Monitor-Analyze-Plan-Execute framework over a shared Knowledge pool (MAPE-K), as depicted in Fig.~\ref{fig:SystemModel}. The model is composed of three key components: (1) a \textit{Workload Processor and Predictor} that preprocess the raw workloads and predicts the future workloads, (2) a \textit{Central Teacher Network} that aims to learn the globally optimal policy, and (3) a \textit{Per-Deployment}\footnote{Please note that the term \textit{Per-Deployment} is inherited the naming conventions module of Kubernetes, which consists a set of pods to run an application workload, usually does not need to maintain state.} \textit{Distributed Student Network} responsible for data collection as well as asynchronous for imitation learning.

\subsection{Workload Processor and Predictor}

The \textit{Workload Processor} is composed of key components including a \textit{Workload Preprocessor}, a  \textit{Load Generator} (e.g. Locust \cite{locust2023}), and a \textit{Historical Database} containing historical workloads. It functions to extract necessary workload data and attributes from the \textit{Historical Database}, preprocesses the dataset via \textit{Workload Preprocessor}, and handles realistic user interactions with the \textit{Load Generator}. It transforms these interactions into requests allocated to the microservice-based cluster. The \textit{Historical Database} stores the historical data derived from realistic traces (e.g. Alibaba traces\cite{guo2019limits}). Detailed dataset information such as timestamps, machine identifiers and different types of resources are contained in these workloads. 

The \textit{Workload Predictor} is a critical component, responsible for forecasting future loads coming into the system. It communicates to the auto-scaler about the required amount of resources to be scaled-in or scaled-out. 
The \textit{Workload Predictor} obtains preprocessed workload data from the \textit{Workload Processor} and employs predictive models such as Long Short-Term Memory (LSTM)\cite{hochreiter1997long} or Gated Recurrent Unit (GRU) to forecast future workloads. The responsibilities of the \textit{Workload Predictor} include managing the training process of the model, updating the trained model when necessary, and deploying the trained models in their final stages. This module interacts with the Workload Analyzer and Workload Generator, receiving historical system data as input.

\subsection{Central Teacher Network}

The use of multi-agent RL often achieves sub-optimal solutions \cite{bucsoniu2010multi}, indicating a need for a central agent (\enquote{teacher}) module that suggests actions to another agent (\enquote{student}) \cite{zimmer2014teacher}. This module investigates the global state, formulating an optimal policy function to enhance system performance. The collected data is employed to train the decentralized student network via the technique of imitation learning \cite{hussein2017imitation}.

Existing performance modelling-based or heuristic-based strategies suffer several limitations, including struggles with model reconstruction and retraining due to their inability to adapt to dynamic system statuses \cite{Cao2020TNSM}. These strategies require considerable expert knowledge \cite{rahman2019predicting}, demanding extensive efforts for the interpretation of microservice workloads and infrastructure. 
Compared with them, RL is suitable for formulating resource provisioning policies due to its capabilities of offering a close feedback loop, exploring scaling actions, and formulating optimal policies independent of erroneous assumptions. This allows for learning directly from actual various workloads and operational conditions, providing deeper insights into how changes in low-level resources influence application performance. We have incorporated two techniques below to achieve our objective:

\subsubsection{Twin Delayed Deep Deterministic Policy Gradient (TD3) \cite{fujimoto2018addressing} }

To overcome the limitation of overestimation of $Q$ value due to the flexible cloud environment, we utilize the TD3 algorithm, an advanced model-free, actor-critic RL framework. Our RL formulation provides two distinct advantages:
\begin{itemize}
\item The model-free RL eliminates the need for precise modelling of the complete distribution of states or the environment dynamics (transitions between states). In the event of microservice updates, the simulations of state transitions used in model-based RL become inefficient. 
\item The actor-critic framework, merging policy-based and value-based methods, is suitable for continuous stochastic environments. This accelerates convergence, reduces variance, and provides a robust and efficient solution for managing the dynamic nature of container-based environments.
\end{itemize}

\subsubsection{Retraining Notifier}

The Retraining Notifier signals when the system requires retraining. There are instances where the imitation learner might not make efficient decisions, triggering the Retraining Notifier and initiating a retraining state. This is typically initiated by the following scenarios:

\begin{itemize}

\item Insufficient exploration by the teacher network: If the teacher network encounters a rarely seen state about which it lacks confidence, it might fail to guide the student appropriately. This can potentially result in the cloud service scaler misjudging the system's scaling needs.

\item Sub-optimal learning state of the student deployment network: In this case, despite the central network learning an optimal policy function, the distributed deployment network fails to establish a correlation between the optimal policy and the domain information it possesses. This can occur if the information provided by the central teacher network is over-complex, requiring an adjustment in the number of epochs that the distributed deployment net runs to achieve convergence.

\item Outdated knowledge held by the student deployment network: For instance, if  the majority of users previously searched for tickets, the student net may over-rely on this pattern. If the request composition changes, such as users predominantly purchasing tickets, the deployment of student nets would require retraining.

\end{itemize}

\color{black}
\subsection{Distributed Student Deployment Network}

The Distributed Student Deployment Network module operates in a distributed fashion, enabling the parallel execution of scaling actions for microservices. Rather than maintaining global states, it relies on local information to function. This module consists of two key components: the \textit{Deployment Buffer}, which stores the policies communicated by the teacher node for training the student network, and the \textit{Imitation Learner}, which scales each deployment independently.

\subsubsection{Deployment Buffer}
\label{deployment buffer}

It maintains the policy transmitted by the Central Module, and stores the state of the deployment when the policy is recorded. This policy-state pairing is utilized to train the Deployment Student Network via imitation learning, which aims to ensure replicas' successful behaviours in a given context.

\subsubsection{Imitation Learner}
\label{Imitation Learner}

The Distributed Student Deployment Net is characterized by its lightweight nature and lower resource requirements compared to the Central Teacher Network. This characteristic facilitates more frequent and adaptable decision-making. The system operates in two distinct modes: learning mode, where it receives guidance from the Central Network, and acting mode, where it autonomously makes decisions based on its state, including factors like resource utilization. This dual-mode configuration enables continuous learning and adaptation, leading to optimized resource allocation and enhanced system performance.
\color{black}


\section{Distributed RL Algorithm for Scaling Microservice}
\label{algorithm_design}
In this section, we introduce the algorithm design of our proposed DRPC approach, which can achieve efficient distributed decisions. 

In DRPC, the RL-based resources provisioning is modelled as a Markov Decision Process, it views the state \( s_t \in S \) as the current microservices system status and interprets the action \( a \in A \) as a scaling operation modifying system status and allocated resources. We use $M = (m^1, m^2,$ \ldots, $m^i)$ to denote the physical machines provisioning resources for the microservices in our system. For each physical machine, say $m^i$, the amount of resources that can be allocated are denoted as $R\textsuperscript{i} = (r^{i,1}, r^{i,2}, \ldots, r^{i,j})$, where the $j$ represents the type of resources, such as CPU and memory. Finally, the set of actions that can be performed on each physical machine is denoted as $A^i = (a^{i,1}, a^{i,2}, \ldots, a^{i,k})$. The actions indicate the amount of resources that can be altered on each machine. The supported actions are \textit{Horizontal\_scaling}, \textit{CPU\_scaling}, \textit{Memory\_scaling}, and \textit{Brownout}. A positive or negative sign before an action implies the addition or reduction of resources to a specific machine, respectively. The term 'Brownout' is a binary indicator for whether a brownout can be triggered on the machine. If this value is set to True, the minimum allowable replicas for horizontal scaling would be set to 0. The collective action space for the system, $A = \prod_{i=1}^{I} \prod_{k=1}^{k} A^{i,k}$, is the product of the action spaces for each machine.\\ 
\color{black}

\subsection{System-wide Action Execution}
\begin{algorithm}[t]
    \color{black}
    \caption{DRPC: System-wide Action Execution} 
    \label{alg:Overalexecution}
    \SetAlgoLined
    \SetKwInOut{Input}{Input}\SetKwInOut{Output}{Output}
    \SetKwFor{ForAll}{forall }{do}{end forall}
    \Input{A central module for controlling scaling actions}
    \Output{None} 
    
    Initialize $trainingMode$ as True or False based on the system state\;
    
    \While{True}{
        \uIf{$trainingMode$}{
            // Stage 1: The central module explores the state and chooses actions
            $actions$ = central module.detection()\;
            centralModule.Scale($actions$) based on Alg. 2\;
            \ForEach{deploymentAction in getActionByDeployment($actions$)}{
                // Training deployment networks
                deployment.train(deploymentAction)\;
            }
        }
        \Else{
            // Stage 2: Deployments make independent scaling decisions
            \ForEach{deployment in allDeployments}{
                deployment.Scale() based on Alg. 2\;
            }
        }
        // Check if the retraining notifier is triggered and update $trainingMode$
        $trainingMode$ = retrainingNotifier()\;
    }    
\end{algorithm}

As illustrated in Algorithm \ref{alg:Overalexecution}, the execution of the system-wide action plan progresses through two distinct stages: an exploration phase and a distributed provision phase.

In Stage 1, the exploration phase, (lines 3-9), the central teacher node explores the state space, choosing multi-dimensional scaling actions (horizontal, CPU, memory scaling, or maintaining the current state) for each deployment as described in the section \ref{sec:Central Module}. Simultaneously, deployment-level networks initiate training.

In Stage 2, the distributed provision phase, (lines 10-12), decision-making transitions to deployment-level networks once the central node's exploration is deemed sufficient, with the same scaling options. This phase continually updates the teacher network's buffer with action-state pairs, as described in section \ref{sec:RetrainingNotifier}. The system alternates between these stages based on the retraining notifier (line 14), enabling the central node to re-explore states or delegate decision-making based on its readiness.
\color{black}

\color{black} Algorithm \ref{alg:GeneralScalingProcedure} illustrates the general scaling procedure of DRPC. The algorithm seeks advice from the Scaling Agent to obtain Q-values that guide resource adjustments (line 1). This includes CPU usage (lines 2-4), and memory utilization (lines 5-7). Such guidance manifests as actions within the DRPC framework. If the recommended scaling for CPU or memory is significant, adjustments are made proportionally to ensure system stability or enhance efficiency. Adding or removing replicas (lines 9-13) follows a similar process as above, except if a microservice is pre-configured to support brownout, the minimum number of replicas can be set to 0.

\subsection{Workload Processor and Predictor}
\label{Workload Generator}
The connection between different levels of resource usage and workloads is determined using a deep neural network model for the workload analysis, as shown in Fig. 2. To mimic realistic resource utilization, we apply data from Alibaba traces, which contain workload traces from 4000 machines, encompassing eight days of resource usage data. 
The performance profiling procedure is as follows: we define a scheduling interval of 5 minutes, over which we gradually increase the number of requests, with each test case comprising 200 requests sent to the host. 
A three-layer Multi-Layer Perceptron (MLP) is applied to the profiling data, thereby efficiently representing this relationship. Consequently, we can translate the host utilization into the number of requests to our system for any given level of utilization. Then, our workload generator produces these requests in accordance with the current user type composition.
\color{black}

We employ multivariate time series forecasting (MTFS), which converts MTFS problem into a supervised learning task. 
The Gated Recurrent Unit (GRU) \cite{GRU}, a type of recurrent neural network (RNN), is then used as the prediction method, as it has been validated with good performance in \cite{Xu2022TOIT} for MTFS. The GRU solves the vanishing gradient problem encountered in conventional RNNs through the utilization of gating mechanisms. 
		
		
		
			
			
			
			
			
			
			
			
			
			
		

\begin{table}[]

		\centering
		\caption{Workloads Prediction Accuracy}
		\label{tab:predictionalg} 
             \resizebox{0.5\textwidth}{!}{
		\begin{tabular}{|c|c|c|c|c|c|}
			\hline
			\textbf{\color{black}Predicted length (mins)\color{black}} & 5     & 10     & 15     & 20     & 25     \\ \hline
			MSE           & 0.0025 & 0.0027 & 0.0033 & 0.0049 & 0.0057 \\ \hline
		\end{tabular}
             }
	\end{table}
\color{black}Table ~\ref{tab:predictionalg} shows the mean square errors (MSE) of actual utilization versus predicted utilization for Alibaba workloads over different predicted length. \color{black}With MSE values ranging between 0.002 and 0.006, it shows that the workload prediction algorithm can achieve good performance in resource utilization prediction.

\subsection{Central Module}
\label{sec:Central Module}
\color{black}

\begin{algorithm}[t]
\SetAlgoLined
\DontPrintSemicolon
\caption{DRPC: General Scaling Procedure}
\label{alg:GeneralScalingProcedure}
\KwIn{Scaling Agent: $Agent$, CPU usage: $Cpu\_usage$, Memory usage: $memory\_usage$, number of replicas: $replicas$, step size for CPU: $Cpu\_step$, step size for memory: $Memory\_step$, Microservice state: $thisMicroservice.Brownout$}
\KwOut{Updated $Cpu\_usage$, $memory\_usage$, $replicas$}
[$Cpu\_Scaling$, $Memory\_Scaling$, $Horizontal\_Scaling$] $\leftarrow Agent.getQvalue()$\\

\If{$|Cpu\_Scaling| > 0.5$}{
    $Cpu\_usage \leftarrow Cpu\_usage + Cpu\_Scaling.Clip(-1, 1) \times Cpu\_step$
}
\If{$|Memory\_Scaling| > 0.5$}{
    $memory\_usage \leftarrow memory\_usage + Memory\_Scaling.Clip(-1, 1) \times Memory\_step$
}
\If{$|Horizontal\_Scaling| > 0.5$}{
    \eIf{$replicas + \lceil Horizontal\_Scaling \rceil < 0$ \text{and} $thisMicroservice.Brownout == False$}{
        exit
    }{
        $replicas \leftarrow replicas + \lceil Horizontal\_Scaling \rceil$
    }
}
\end{algorithm}

Conventional RL models such as Sarsa \cite{ChainsFormer2023} and Q-learning encounter challenges when dealing with infinite or continuous states and actions, a common scenario in microservices characterized by variable CPU and memory usage. To address this issue, we employ the TD3 algorithm, a more sophisticated approach well-suited for continuous, high-dimensional spaces. TD3 utilizes twin critic networks to reduce bias, incorporates delayed policy updates for stable learning, and employs target policy smoothing to prevent overfitting. These attributes render TD3 robust and efficient, making it an ideal choice for optimizing autoscaling in dynamic microservices environments.
\color{black}

\subsubsection{TD3 modelling}
Our dual objectives are improving the Quality of Service (QoS) and maximizing the utilization of physical machines. Our reward model, encompassing response time $R_{qos}(rt)$ 
and resource utilization $R_{util}(u)$ are accordingly formulated in Equations (\ref{eq:reward}) and (\ref{eq:rewardutil}):\\

	\begin{equation}
		\label{eq:reward}
		R_{qos}(rt) = \begin{cases}
			e^-({\frac{rt-RT_{max}}{RT_{max}})^2} &, rt >  RT_{max} \\
			1 &, rt \leq RT_{max},
		\end{cases}
	\end{equation}

where the maximum tolerant latency, $RT_{max}$, is pre-defined into the QoS reward function. Normal system operation rewards 1, while performance that exceeds $RT_{max}$ is penalized, gradually approaching 0, thus discouraging SLO violation.\\
\color{black}

Resource utilization $R_{util}(u)$, measured using a devised model shown in Equation~(\ref{eq:rewardutil}), 
\begin{equation}
	\label{eq:rewardutil}
	R_{util}(u) = \begin{cases}
		\frac{\sum_{k=1}^K \sum_{r=1}^R (U_{r,k}^{pred}-u_{r,k})^3}{K} + 1 &, u_{rk} \leq U_{r,k}^{pred} \\
		\frac{\sum_{k=1}^K \sum_{r=1}^R (u_{r,k} - U_{r,k}^{pred})^3}{K} + 1 &, u_{r,k} > U_{r,k}^{pred},
	\end{cases}
\end{equation}
which guides the system towards resource conservation. Here, $K$ and $R$ represent the $kth$ physical machine and $rth$ resource type residual on the $kth$ physical machine, respectively, for example, the $u_{1,1}$ indicate the 1st adjustable resources on the 1st machine. $U_{r,k}^{pred}$, the predefined utility, represents the ideal utility for a certain resource type $r$ on machine $k$. Proximity to this ideal utilization is rewarded; under-provisioning or wastage is discouraged.\\

The final reward value as shown in Equation (\ref{eq:finalreward}) is a combination of both response time and resource utilization, The objective is:
\begin{equation}
	\label{eq:finalreward}
	r(s_t, a_t) = \frac{R_{qos}(rt)}{R_{util}(u)}.
\end{equation}

\color{black}
The final objective is to decrease the response time while keeping the system running in a stable state  $U^{pred}$ as shown in Equation (\ref{eq:obj}).\\
\begin{equation}
\label{eq:obj}
\min_{u \in Total\_Resources} | U^{pred}- R_{\text{util}}(u)|    \land  \min_{rt \in RT} R_{qos}(rt).
\end{equation}

\color{black}

\subsubsection{TD3 Implementation Details}

\begin{algorithm}[t]
\caption{DRPC: TD3 Algorithm}
\label{alg:TD3}
\DontPrintSemicolon
\SetAlgoLined
\KwResult{TD3 (Twin Delayed Deep Deterministic policy gradient)}

Initialize actor network $\pi$ and critic networks $Q_{\theta_1}$, $Q_{\theta_2}$ with random parameters $\theta, \theta_1, \theta_2$\;
Initialize target networks $\pi'$ and $Q'$ with weights $\pi' \gets \pi, Q'_{\theta_{1'}} \gets Q_{\theta_1}, Q'_{\theta_{2'}} \gets Q_{\theta_2}$\;
Initialize replay buffer $R$\;
\For{episode = 1 to M}{
    Observe state $s$\;
    \For{t = 1 to T}{
        Select action $a = \text{clip}(\pi(s) + \epsilon , a_{High}, a_{Low}) \text{ where }\epsilon \sim \mathcal{N}$ and execute it\;
        Observe next state $s'$, reward $r$ and done signal $d$ to end episode\;
        Store $(s, a, r, s', d)$ in $R$\;
        \If{it is time to update}{
            Sample a batch B of transitions $(s, a, r, s', d)$ from $R$\;
            $a' \gets \text{clip} (\pi'(s') + \text{clip}(\epsilon, -c, c), a_{High}, a_{Low}) $\;
            $y \gets r + \gamma (1 - d) \min_{i=1,2} Q'_{\theta_{i'}}(s', a')$\;
            Update $Q_{\theta_i}$ by minimizing the loss: $L(\theta_i) = \frac{1}{|B|}\sum_{(s, a, r, s', d) \in B}(Q_{\theta_i}(s, a) - y)^2$\;
            \If{$t \mod \text{policy\_update} == 0$}{
                Update $\pi$ by one step gradient ascent using the loss: $L(\theta) = \frac{1}{|B|}\sum_{s \in B} -Q_{\theta_1}(s, \pi(s))$\;
                Soft update the target networks: $\theta_{i'} \gets \rho \theta_{i'} + (1 - \rho)\theta_i$ (for both actor and critic)\;
            }
        }
        $s \gets s'$\;
    }
}

\end{algorithm}
 \color{black}
 
As shown in Algorithm \ref{alg:TD3}, TD3 begins by initializing the actor and critic networks, as well as the replay buffer (lines 1-3). It then selects an action from the actor network, executes the action, and observes the subsequent state, reward, and episode termination status (lines 4-8). This tuple $(s, a, r, s', d)$ is stored in the replay buffer (line 9). Once enough data has accumulated and it is time for an update, TD3 samples a batch of transitions from the buffer, which includes data from both the central and distributed modules (lines 10-11).
In certain instances, scale-out may considerably diminish the SLO violation, while scale-in, by circumventing the communication time, could realize superior response times after several scaling operations. The robust reward resulting from scale-out might engender a fragile, or erroneous, "sharp peak," potentially obscuring our model's capacity to investigate the long-term impact of Scale-in. Furthermore, due to the dynamic nature of cloud environments, where internet connections fluctuate, some reallocation actions may reduce QoS violations more effectively than anticipated. Ultimately, this can lead to an overestimation of Q values.
To alleviate this, a clipped (limits values to a range) \ colour {black} noise, between bound $-c \text{ and } c$, $\epsilon$ is incorporated into the target policy $\pi'(s')$ to get a final policy $a'$ and that policy is ensured to lie within $a_{Low} \text{ and } a_{High}$ as shown in Equation (\ref{eq:clip}) (line 12).

\begin{equation}
\label{eq:clip}
a' \gets \text{clip} (\pi'(s') + \text{clip}(\epsilon, -c, c), a_{High}, a_{Low}) ,
\end{equation}
where both Q-functions employ a singular target. Moreover, as microservices are deployed in a distributed manner and communicate via cable, certain scaling actions may lead to a significant reduction in communication time due to internet fluctuations, ultimately generating a substantial reward. To circumvent the overestimation of rewards, Clipped double-Q learning is employed. Both Q-functions utilize a singular target $y(r,s',d)$, calculated by summing the immediate reward $r$ and the minimum value obtained from the two Q-functions $Q'_{\theta_{i'}}(s', a')$, multiplied by the discount factor $\gamma$. If the subsequent state is a terminal state ($d = 1$), no future reward is considered (line 13) as shown in Equation (\ref{eq:future_reward}):
\begin{equation}
\label{eq:future_reward}
 y(r,s',d) \gets r + \gamma (1 - d) \min_{i=1,2} Q'_{\theta_{i'}}(s', a').
\end{equation}

The loss indicated as $L(\theta_i)$ in line 14, for the $i_{th}$ critic network, is calculated as the average mean of the sum of squared differences between the target Q-value $y$, and the critic network predicted Q-value $Q_{\theta_i}(s, a)$, across a mini-batch experiences $B$ containing a set of $(s, a, r, s', d)$ experience. The number of such experiences is denoted by $|B|$ as formulated in Equation (\ref{eq:B}):

\begin{equation}
\label{eq:B}
L(\theta_i) = \frac{1}{|B|}\sum_{(s, a, r, s', d) \in B}(Q_{\theta_i}(s, a) - y)^2.
\end{equation}

To minimize the variability in microservice communication and to learn a more stable $Q$ function, the policy will only be updated by one step of gradient ascent once every $policy\_update$ (line 15) times using the loss function (line 16) with Equation (\ref{eq:update}):

\begin{equation}
\label{eq:update}
L(\theta) = \frac{1}{|B|}\sum_{s \in B} -Q_{\theta_1}(s, \pi(s)).
\end{equation}

Finally, the target network will be updated (lines 17-20).

\subsection{Deployment Unit with Imitation Learning}

Imitation learning involves training an agent to perform tasks by emulating the actions of an expert. This approach has gained popularity due to its capability to teach complex behaviors without the need for explicit programming or heavy reliance on reward signals, which is common in traditional RL. In our scenario, the deployment unit utilizes its information to imitate the resource adjustment actions of the Central module. This objective, as shown in Eq. (\ref{eq:MSE}),  is accomplished by minimizing the mean square error between the Q-values $Q_{\text{Ctr}}(s)$ generated by the central network and those $Q_{\text{Dtr}}(s)$ produced by the distributed network under the same state $s$.

\begin{equation}
\label{eq:MSE}
\min \left( Q_{\text{Ctr}}(s) - Q_{\text{Dtr}}(s) \right)^2
\end{equation}

When the retraining notifier signals deployment readiness, scaling actions, guided by human knowledge-driven frequency, are executed simultaneously and asynchronously. State-action-reward pairs are then gathered, combined with data from other deployments, and utilized to fine-tune the policy of the central network.
\color{black}
\subsection{Retraining Notifier}
\label{sec:RetrainingNotifier}
When the RL-based model needs to be re-trained based on the conditions as introduced in Section IV.B, the \textit{Retraining Notifier} as shown in Algorithm \ref{alg:retrainNotifier} informs the teacher network to repeat the training phase whenever the student network faces insufficient information or the average reward from recent actions falls below a predetermined threshold. It initializes an array, $rewardHistory$, to record the last $npr$ rewards and the iterator number $iter$; continuously, the algorithm calculates the current action's reward, storing it cyclically in $rewardHistory$ (lines 1-7). If collected rewards are below the defined number $npr$, retraining will not be triggered due to insufficient data (line 8). However, if the average of $rewardHistory$ drops below the threshold $TH$, the retraining process will be triggered, indicating suboptimal performance 
 (lines 10-12). Otherwise, the system continues its distributional operations (lines 13-15).

\begin{algorithm}[t]
    \color{black}
    \caption{DRPC: Retraining Notifier} 
    \label{alg:retrainNotifier}
    \SetAlgoLined
    \SetKwInOut{Input}{Input}\SetKwInOut{Output}{Output}
    \Input{Retraining Threshold $TH$, number of past rewards to monitor $npr$, action sets $a_t=\{a_k^i(t)|i\in \{0, 1, \ldots, I\},$ $k\in \{1, 2, \ldots,K\}\}$}
    \Output{Boolean indicating whether retraining mode is triggered} 
    
    Initialize an array $rewardHistory$ of size $npr$ to store the rewards from past actions\;
    
    Initialize an Iterator $iter$ = 0\;
    
    \While{True}{
        Calculate the current action reward $currentReward$ using get\_reward($a_t$)\;
        
        Store the current reward at the position $iter$ \% $npr$ in $rewardHistory$\;
        
        $rewardHistory[iter \% npr]$ = $currentReward$\;
        
        $iter$ ++\;
        
        \uIf {$iter < npr$}{
            // Not enough data gathered yet, do not trigger retraining
            return False\;            
        }
        \uIf {average($rewardHistory$) $<$ $TH$}{
            // The average reward is below the threshold, trigger retraining
            return True\;
        }
        \Else{
            // The average reward is above the threshold, do not trigger retraining
            return False\;
        }
    }    
\end{algorithm}


\section{PERFORMANCE EVALUATIONS}
\label{performance}
\begin{figure*}[t!]
		\centering
		\begin{subfigure}{0.24\linewidth}
			\includegraphics[width=0.99\linewidth]{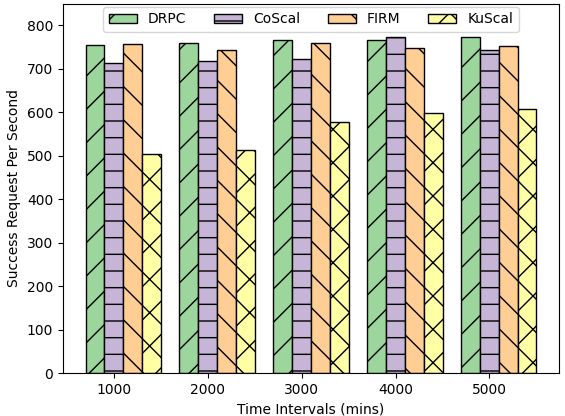}
            \caption{Number of success requests}
            \label{fig:SuccRequestPerSecond}
		\end{subfigure}
      \begin{subfigure}{0.24\linewidth}
			\centering
			\includegraphics[width=0.99\linewidth]{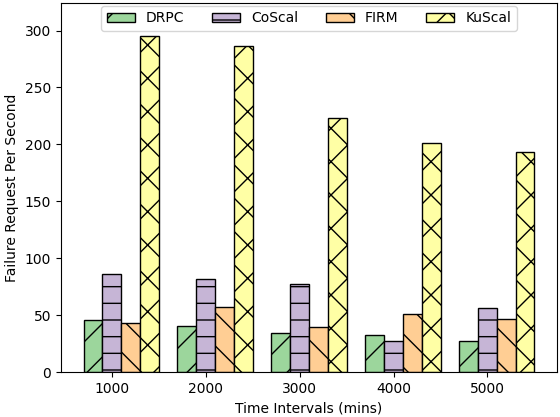}
			\caption{Number of failure requests}
			\label{fig:FaliureRate}
		\end{subfigure}
		\begin{subfigure}{0.24\linewidth}
			\centering
			\includegraphics[width=0.99\linewidth]{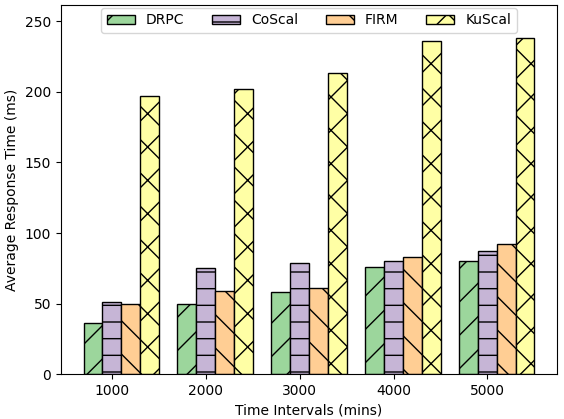}
            \caption{Average response time}
           \label{fig:avg_time}
		\end{subfigure}
        \begin{subfigure}{0.24\linewidth}
			\centering
			\includegraphics[width=0.99\linewidth]{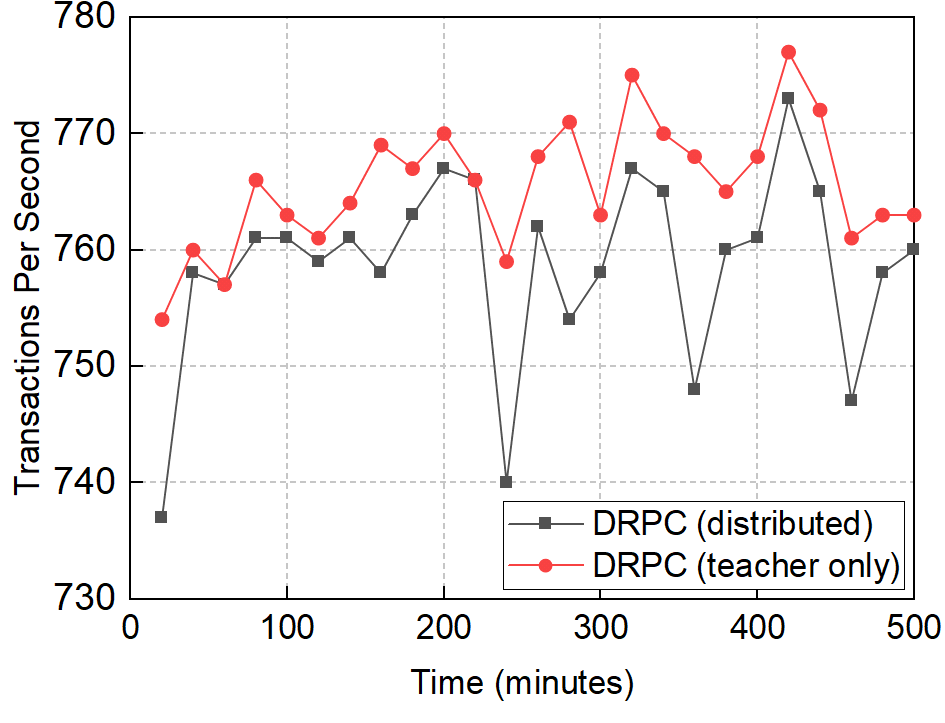}
			\caption{Transactions per second}
			\label{fig:scalability_vs_self}
		\end{subfigure}
		\caption{Performance comparison of KuScal, FIRM, CoScal, and DRPC}
		\label{fig:avg_parameters}
	\end{figure*}

To assess the efficacy of DRPC in the autoscaling of microservices, we conduct experiments on a realistic testbed based on Kubernetes. We detail the experimental settings, benchmarks, and a comprehensive analysis of the results. The primary objective is to verify that a distributed asynchronous framework can improve performance, potentially offering potential directions for future research.\color{black}

\subsection{Experimental Setup}

We utilize TrainTicket as a testbed in our motivational example in Section \ref{sec:motivation}. It provides functionalities such as ticket purchases, availability checks, cancellations, and news browsing, simulating backend processes like verification code generation, user login checks, and database ticket availability searches. The platform establishes latency-measuring chains using domain name service for service connectivity. Load balancers or service routers based on cloud server tests are employed. Each TrainTicket microservice has its own database, reducing wait times and facilitating scheduling for developers.

We use a cluster with five machines (one master and four workers) for microservice-based cluster. Each machine has 8 CPU and 8GB memory. This prototype system is implemented using a suite of toolkits, including Cgroup v2, Python, TensorFlow, PyTorch, Sklearn, and Locust, which support deep learning and reinforcement learning environments.

We simulate four user types (Normal Query Users, Ticket Buyers, Cancel Ticket Users, and Admin Users) with various ratios in our cluster using the Alibaba work trace, each assigned with a unique ID for tracking. This emulates real-world user behavior dynamics, where service requests fluctuate over time. Data is collected every 200 milliseconds and averaged over 1000-minute intervals in accordance with the existing work \cite{xu2022CoScal}\cite{qiu2020firm}.

\color{black}




\subsection{Baselines}
Several state-of-the-art approaches from both industry and academia have been selected as baselines. 

\textbf{KuScal} \cite{burns2019kubernetes} is a Kubernetes mechanism using horizontal scaling to adjust replicas dynamically. It determines replica quantity based on resource fluctuations, adjusting microservice numbers via continuous real-time resource utilization tracking and predefined CPU utilization thresholds. The threshold is set to be 75\% as it is evaluated as the most effective in some existing work surveyed in  \cite{carrion2022kubernetes}.

\textbf{CoScal} \cite{xu2022CoScal} divides resource usage into quarterlies, applies an approximated Q-learning algorithm to devise a policy, and supports horizontal, vertical scaling, and brownout. Using the GRU unit for workload prediction, it forecasts workload, consulting a trained table to guide pod-level scaling actions, and even optimizing non-critical path pods.

\textbf{FIRM} \cite{qiu2020firm} systematically allocates cloud resources by identifying the critical microservice path and using machine learning to target specific instances needing optimization. The process strategically moves from the longest to the shortest chain, prioritizing longer chains to reduce microservice time delays and boost efficiency.

\subsection{Experiment Analysis}

	\begin{figure*}[ht!]
		\centering		
      \begin{subfigure}{0.32\linewidth}
			\centering
			\includegraphics[width=0.9\linewidth]{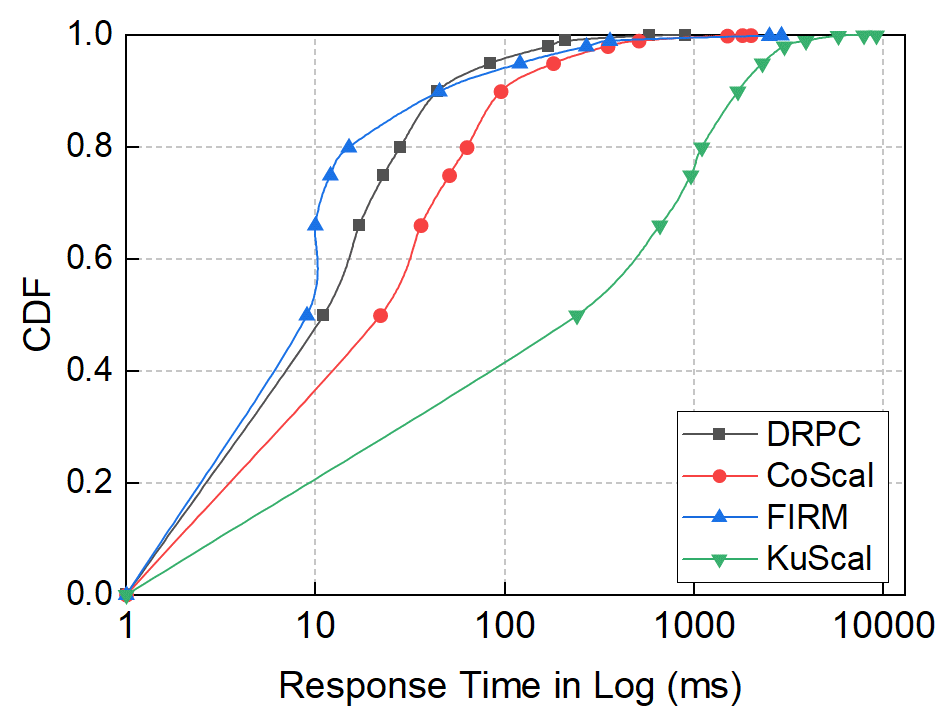}
			\caption{CDF of response time}
			\label{fig:CDF}
		\end{subfigure}
  \begin{subfigure}{0.3\linewidth}
			\centering
			\includegraphics[width=0.9\linewidth]{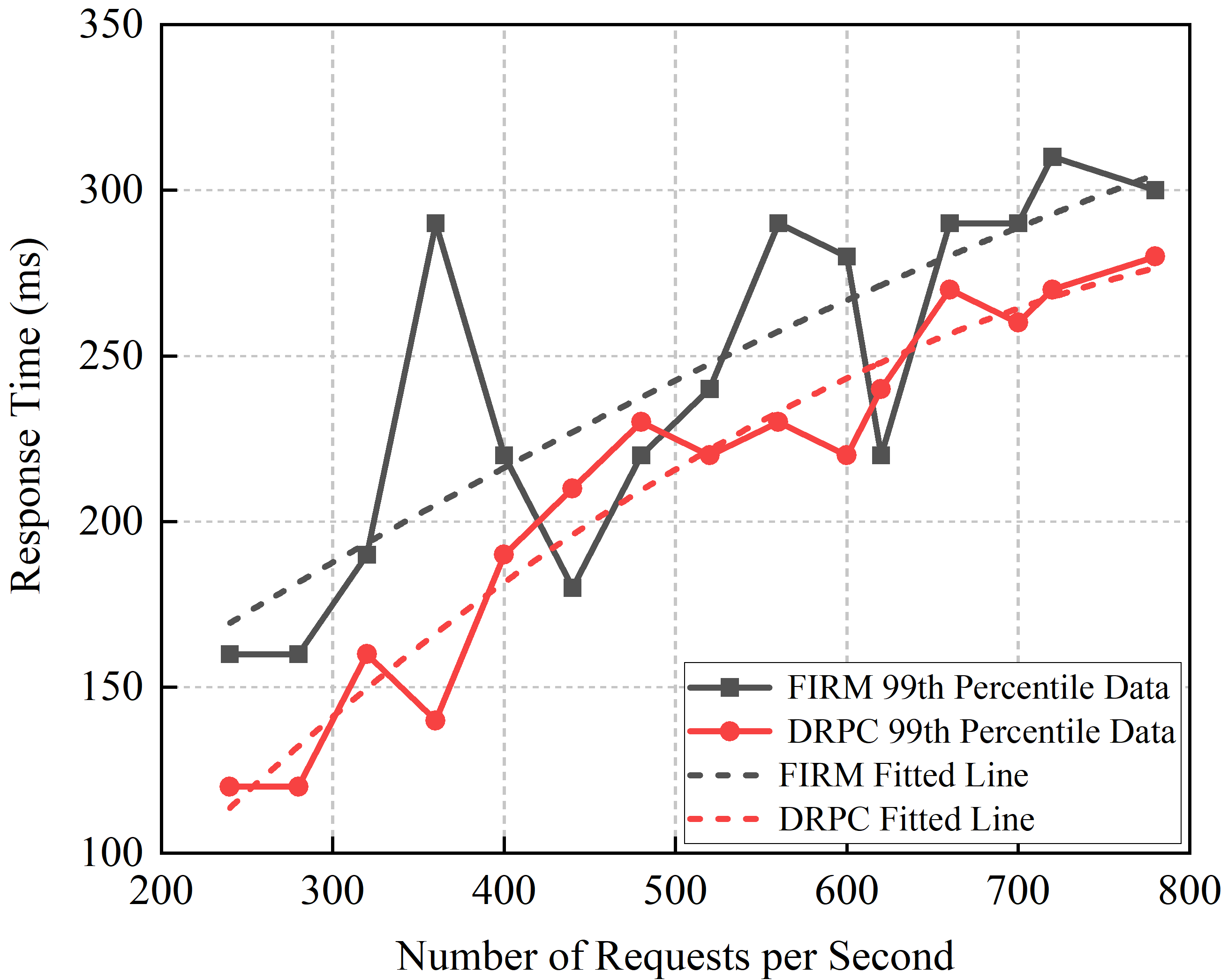}
            \caption{Scalability comparison}
           \label{fig:scalability_vs_firm}
		\end{subfigure}
  		\begin{subfigure}{0.32\linewidth}
			\centering
			\includegraphics[width=0.9\linewidth]{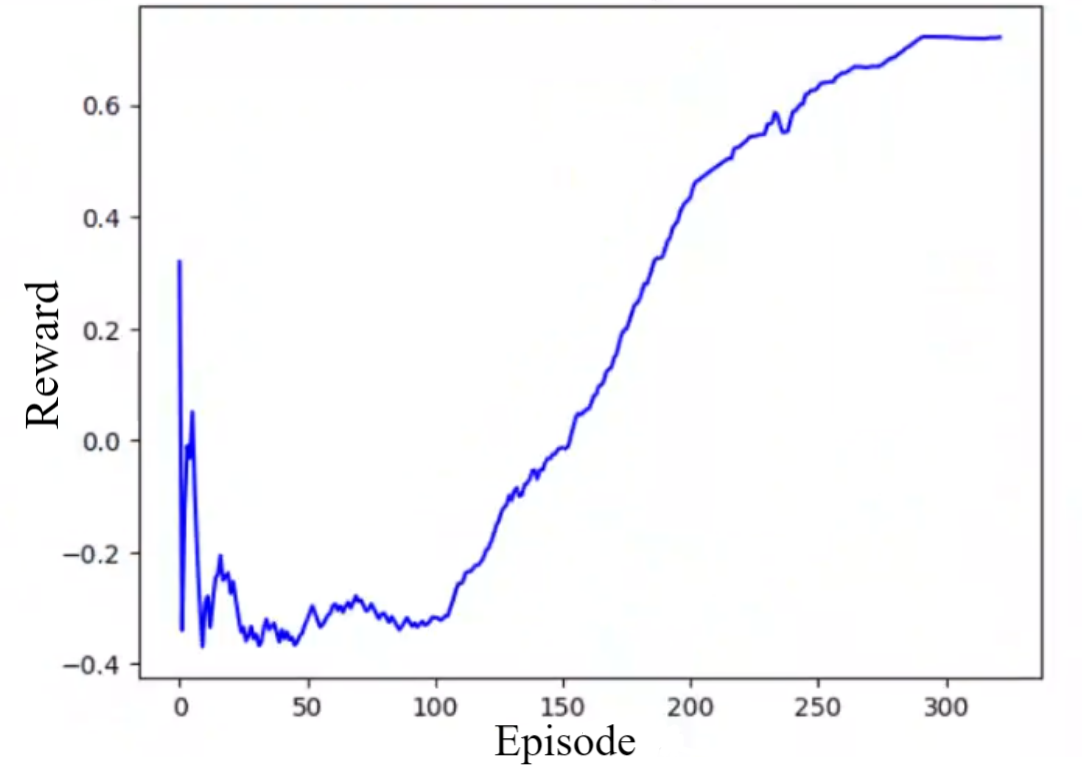}
            \caption{Reward convergence of DRPC}
           \label{fig:centralReward}
		\end{subfigure}
		\label{fig:cdf_and_reward}
  		\caption{CDF comparison and reward convergence}

	\end{figure*}
\color{black}
Several widely used metrics, including the number of success requests, failure rate and response time have been utilized to evaluate the performance of our proposed approach and the baselines.

\color{black}

\subsubsection{Computation Efficiency Analysis}




Based on the efficient design of our distributed student network, \color{black}DRPC necessitates considerably fewer computational resources compared to FIRM
, with trainable parameters being 643 for our model and 11,352 for FIRM. DRPC utilizes only $6 \times 10^{-5}$ of CPU time to determine the subsequent action, compared to $4 \times 10^{-3}$ taken by FIRM. Our model also showcases superior efficiency, with a peak CPU utilization of 0.8 as opposed to FIRM's maximum usage of 2.0. This illustrates that our approach is markedly more resource-efficient than FIRM. \color{black}
DRPC incorporates 40 services, and enables simultaneous application of CPU, memory, and horizontal scaling for each deployment, executing up to 120 actions per time step. KuScal ranks second, when CPU utilization for all services surpasses the threshold, and all services undergo horizontal scaling. 
However, FIRM can only scale one resource type per step, as FIRM's Actor-critic network estimates the Q-value for scaling actions and applies the action with the highest Q-value. 

\color{black}The offline training phase required approximately 10 days to converge, with the majority of this time dedicated to the data collection process within the cluster. However, once the initial training is completed, the model can be fine-tuned within minutes via online learning to adapt to changes in the composition of user requests, which can handle the significant changes in realistic system. \color{black}

\subsubsection{Comparison of the Number of Success Requests per Second }


Fig. ~\ref{fig:SuccRequestPerSecond} presents the requests per second processed by four distinct algorithms. We partitioned the 5000-minute data into five time periods, with each period representing the average results over the corresponding period. 
We observed that, across the five time periods, FIRM is capable of processing a higher number of requests compared to our method during the initial periods. However, our method outperformed FIRM over the subsequent four periods. Likewise, CoScal also demonstrated the ability to adapt and improve over time, but it did not outperform our method or FIRM consistently. Furthermore, the difference between our method, CoScal, and FIRM was relatively small during the initial periods, until our method started surpassing them. 
It can be also reasoned that a scheduling algorithm solely reliant on the longest chain becomes increasingly challenging to handle workloads under significant variances in request rate over time. This might be caused by the longest chain typically signifying more requests at the onset of the scenario when chain variations are minimal. However, as the load changes over time, the chains experiencing high loads, which may not necessarily be the longest chains, become the actual targets for optimization. 
In summary, FIRM and Coscal have achieved good performance in handling requests successfully. Compared to FIRM, DRPC improves the success rate of requests by around 2\%.

\subsubsection{Comparison of Failure Rate}



\color{black}
As shown in Fig. \ref{fig:FaliureRate}, DRPC has a failure rate of 4.5\%, which is notably lower than FIRM's 5.9\%, CoScal's 8.2\%, and Kuscal's significantly high rate with 29\%. When we evaluate the improvement of DRPC over the other methods, it reduced the failure requests by 24\% compared to FIRM, 44\% in relation to CoScal, and exhibited an impressive 85\% reduction when compared to KuScal. In summary, DRPC demonstrates a significant performance improvement over the benchmarks. This enhancement is attributed to our design's ability to adjust deployments more frequently, thereby reducing the likelihood of failures compared to other methods.
\color{black}

\subsubsection{Comparison of Response Time}



Fig.~\ref{fig:avg_time} demonstrates the average response time for the four algorithms under the same configurations. 
KuScal shows the longest average response time across all five periods, being three to four times longer than others, as its inability to predict forthcoming requests inhibits it from conducting pre-emptive horizontal scaling to accommodate these incoming requests. 
In contrast, CoScal performs better than KuScal but trails behind FIRM and DRPC. The CoScal model demonstrates an average response time that is generally similar to that of FIRM, except for the last two time periods, where FIRM shows a slight gap.
\color{black}We observed that the DRPC achieves the shortest response time, typically ranging between 40 ms and 70 ms, for most of the time intervals, thereby reducing the average response time by around 15\% compared to FIRM and 24\% compared to Coscal. Remarkably, it reduces the response time by 72.4\% compared to Kuscal.\color{black}

\subsubsection{Scalability Analysis} 
In comparing the transactions per second (TPS) of distributed networks with a centralized teacher network (as shown in Figure ~\ref{fig:scalability_vs_self}), it was observed that asynchronous, frequently updated distributed models consistently surpass the teacher model (centralized) in performance. These models not only emulate the teacher network's patterns but also demonstrate reduced volatility.
\color{black}In Figure ~\ref{fig:scalability_vs_firm}, an analysis shows that as request numbers significantly increase, our method outperforms FIRM with 30\% improvement in terms of response time in average. Additionally, DRPC's lower quadratic coefficient in fitted line than FIRM confirms its effectiveness, scalability, and stability.\color{black}

\subsubsection{Cumulative Distribution Function (CDF) of Response Time Analysis}



\begin{table}
\renewcommand{\thetable}{4}
\caption{Response time at different percentile}
\label{tab:CDF}
\begin{center}
\scriptsize
\begin{tabular}{|c|c|c|c|c|c|c|c|c|}
\hline
 & 50th & 66th & 75th & 80th & 90th & 95th & 99th & 99.99th  \\
\hline
FIRM & \textbf{9} & \textbf{10} & \textbf{12} & \textbf{15} & 45 & 120 & 360 & 2900\\
\hline
CoScal & 22 & 36 & 51 & 63 & 95 & 180 & 510 & 1800 \\
\hline
DRPC & 11 & 17 & 23 & 28 & \textbf{44} & \textbf{84} & \textbf{210}  & \textbf{900}  \\
\hline
Kuscal & 240 & 660 & 960 & 1100 & 1700 & 2300 & 3900 & 7900 \\
\hline
\end{tabular}
\end{center}

 \end{table}

We also utilize the CDF in Table \ref{tab:CDF} to highlight the differences between different approaches, as well as the results in Figure \ref{fig:CDF}.
Until the 90th percentile, the CDF curve of FIRM consistently outperforms DRPC. This difference appears because FIRM allocates requests to specific nodes rather than central ones. Our approach, which releases resources on non-critical chains or nodes, introduces delays on these nodes. \color{black}Since FIRM does not allocate resources to these nodes, resulting in over-provisioning, it handles certain simple requests more quickly than our method. However, beyond the 90th percentile, our strategy efficiently redistributes the released resources to nodes critical for QoS enhancement. In summary, DRPC significantly reduces tail latency by reallocating resources to nodes crucial for enhancing QoS, outperforming baselines consistently.\color{black}
\color{black}

\subsubsection{Reinforcement Learning Convergence Analysis}
As illustrated in Figure \ref{fig:centralReward}, the system requires approximately 300 episodes to reach convergence. Beyond this point, the reward per action experiences no significant increase, prompting us to halt the operation. The reward coverage settles at approximately 0.7, as seen from the figure. This indicates that CPU, memory usage, and latency are approaching an optimal configuration by analyzing the reward function. In terms of actual performance, memory utilization is around 75\%, while CPU utilization is approximately 60\%. The QoS is guaranteed, ensuring that 99\% of requests can be handled within 210 ms.

\color{black}
\section{Conclusions and Future Work}
\label{conclusion}

In this work, we introduced a novel framework for distributed resource provisioning, named DRPC. By adopting an asynchronous, parallel, and differential approach, DRPC facilitates the global optimal allocation of resources. This accommodates the dynamic nature of microservice-based clusters while ensuring QoS. Notably, DRPC incorporates DL methodologies for workload prediction, achieving a higher level of accuracy compared to conventional gradient-based methods. Additionally, it leverages a distributed RL algorithm to make informed decisions on scaling strategies, effectively managing the infinite action-states space associated with microservices. The results, based on realistic testing and comparisons with state-of-the-art algorithms, demonstrate that DRPC outperforms the baselines in terms of successful request rate and average response time, particularly under significantly increased requests.


\color{black}The limitation of this approach is that it increases network usage due to the essential communication of distributed RL. Future research will focus on automating the asynchronous updating process, which currently requires manual setting of timing intervals for specific microservices, to improve system efficiency. \color{black} In addition, we would like to explore anomaly-aware workloads management
under container-based environment, and incorporate our approach into large-scale and production environment (e.g. Alibaba Cloud) with further validations.\color{black}
\color{black}
\section*{SOFTWARE AVAILABILITY}
The codes have been open-sourced to \url{https://github.com/vincent-haoy/DRPC} for research usage.


\bibliographystyle{IEEEtran}

\bibliography{references}

\begin{thebibliography}{10}
\providecommand{\url}[1]{#1}
\csname url@samestyle\endcsname
\providecommand{\newblock}{\relax}
\providecommand{\bibinfo}[2]{#2}
\providecommand{\BIBentrySTDinterwordspacing}{\spaceskip=0pt\relax}
\providecommand{\BIBentryALTinterwordstretchfactor}{4}
\providecommand{\BIBentryALTinterwordspacing}{\spaceskip=\fontdimen2\font plus
\BIBentryALTinterwordstretchfactor\fontdimen3\font minus
  \fontdimen4\font\relax}
\providecommand{\BIBforeignlanguage}[2]{{%
\expandafter\ifx\csname l@#1\endcsname\relax
\typeout{** WARNING: IEEEtran.bst: No hyphenation pattern has been}%
\typeout{** loaded for the language `#1'. Using the pattern for}%
\typeout{** the default language instead.}%
\else
\language=\csname l@#1\endcsname
\fi
#2}}
\providecommand{\BIBdecl}{\relax}
\BIBdecl

\bibitem{Newman_2021}
S.~Newman, \emph{Building microservices: Designing fine-grained systems}.\hskip
  1em plus 0.5em minus 0.4em\relax O’Reilly, 2021.

\bibitem{GiovanniTSC2024}
G.~Quattrocchi, D.~Cocco, S.~Staffa, A.~Margara, and G.~Cugola, ``Cromlech:
  Semi-automated monolith decomposition into microservices,'' \emph{IEEE
  Transactions on Services Computing}, vol.~17, no.~2, pp. 466--481, 2024.

\bibitem{Microsoft2021}
O.~M.~A. Khan and A.~Chandaka, \emph{Developing Microservices Architecture on
  Microsoft Azure with Open Source Technologies}.\hskip 1em plus 0.5em minus
  0.4em\relax Microsoft Press, 2021.

\bibitem{SPE2024}
M.~Xu, L.~Yang, Y.~Wang, C.~Gao, L.~Wen, G.~Xu, L.~Zhang, K.~Ye, and C.~Xu,
  ``Practice of alibaba cloud on elastic resource provisioning for large-scale
  microservices cluster,'' \emph{Software: Practice and Experience}, vol.~54,
  no.~1, pp. 39--57, 2024.

\bibitem{pallewatta2023placement}
S.~Pallewatta, V.~Kostakos, and R.~Buyya, ``Placement of microservices-based
  iot applications in fog computing: A taxonomy and future directions,''
  \emph{ACM Computing Surveys}, 2023.

\bibitem{jawaddi2022review}
S.~N.~A. Jawaddi, M.~H. Johari, and A.~Ismail, ``A review of microservices
  autoscaling with formal verification perspective,'' \emph{Software: Practice
  and Experience}, vol.~52, no.~11, pp. 2476--2495, 2022.

\bibitem{XieTSC2024}
S.~Xie, J.~Wang, B.~Li, Z.~Zhang, D.~Li, and P.~C.~K. Hung, ``Pbscaler: A
  bottleneck-aware autoscaling framework for microservice-based applications,''
  \emph{IEEE Transactions on Services Computing}, vol.~17, no.~2, pp. 604--616,
  2024.

\bibitem{kalavri2018three}
V.~Kalavri, J.~Liagouris, M.~Hoffmann, D.~Dimitrova, M.~Forshaw, and T.~Roscoe,
  ``Three steps is all you need: fast, accurate, automatic scaling decisions
  for distributed streaming dataflows,'' in \emph{13th USENIX Symposium on
  Operating Systems Design and Implementation (OSDI 18)}, 2018, pp. 783--798.

\bibitem{luo2023optimizing}
S.~Luo, C.~Lin, K.~Ye, G.~Xu, L.~Zhang, G.~Yang, H.~Xu, and C.~Xu, ``Optimizing
  resource management for shared microservices: A scalable system design,''
  \emph{ACM Transactions on Computer Systems}, 2023.

\bibitem{padakandla2021survey}
S.~Padakandla, ``A survey of reinforcement learning algorithms for dynamically
  varying environments,'' \emph{ACM Computing Surveys (CSUR)}, vol.~54, no.~6,
  pp. 1--25, 2021.

\bibitem{verma2015large}
A.~Verma, L.~Pedrosa, M.~Korupolu, D.~Oppenheimer, E.~Tune, and J.~Wilkes,
  ``Large-scale cluster management at google with borg,'' in \emph{Proceedings
  of the Tenth European Conference on Computer Systems}.\hskip 1em plus 0.5em
  minus 0.4em\relax ACM, 2015.

\bibitem{he2019re}
X.~He, Z.~Tu, X.~Xu, and Z.~Wang, ``Re-deploying microservices in edge and
  cloud environment for the optimization of user-perceived service quality,''
  in \emph{International Conference on Service-Oriented Computing}.\hskip 1em
  plus 0.5em minus 0.4em\relax Springer, 2019, pp. 555--560.

\bibitem{burns2019kubernetes}
B.~Burns, J.~Beda, and K.~Hightower, \emph{Kubernetes: up and running: dive
  into the future of infrastructure}.\hskip 1em plus 0.5em minus 0.4em\relax
  O'Reilly Media, 2019.

\bibitem{Kannan2019}
R.~S. Kannan, L.~Subramanian, A.~Raju, J.~Ahn, J.~Mars, and L.~Tang,
  ``Grandslam: Guaranteeing slas for jobs in microservices execution
  frameworks,'' in \emph{Proceedings of the Fourteenth EuroSys Conference
  2019}, ser. EuroSys '19.\hskip 1em plus 0.5em minus 0.4em\relax New York, NY,
  USA: Association for Computing Machinery, 2019.

\bibitem{hou2020ant}
X.~Hou, C.~Li, J.~Liu, L.~Zhang, Y.~Hu, and M.~Guo, ``Ant-man: towards agile
  power management in the microservice era,'' in \emph{2020 SC20: International
  Conference for High Performance Computing, Networking, Storage and Analysis
  (SC)}.\hskip 1em plus 0.5em minus 0.4em\relax IEEE Computer Society, 2020,
  pp. 1098--1111.

\bibitem{Yu2019}
G.~{Yu}, P.~{Chen}, and Z.~{Zheng}, ``Microscaler: Automatic scaling for
  microservices with an online learning approach,'' in \emph{2019 IEEE
  International Conference on Web Services (ICWS)}, July 2019, pp. 68--75.

\bibitem{liu2019qos}
L.~Liu, ``Qos-aware machine learning-based multiple resources scheduling for
  microservices in cloud environment,'' \emph{arXiv preprint arXiv:1911.13208},
  2019.

\bibitem{gan2019seer}
Y.~Gan, Y.~Zhang, K.~Hu, D.~Cheng, Y.~He, M.~Pancholi, and C.~Delimitrou,
  ``Seer: Leveraging big data to navigate the complexity of performance
  debugging in cloud microservices,'' in \emph{Proceedings of the twenty-fourth
  international conference on architectural support for programming languages
  and operating systems}, 2019, pp. 19--33.

\bibitem{rzadca2020autopilot}
K.~Rzadca, P.~Findeisen, J.~Swiderski, P.~Zych, P.~Broniek, J.~Kusmierek,
  P.~Nowak, B.~Strack, P.~Witusowski, S.~Hand \emph{et~al.}, ``Autopilot:
  workload autoscaling at google,'' in \emph{Proceedings of the Fifteenth
  European Conference on Computer Systems}, 2020, pp. 1--16.

\bibitem{xu2022CoScal}
M.~Xu, C.~Song, S.~Ilager, S.~S. Gill, J.~Zhao, K.~Ye, and C.~Xu, ``Coscal:
  Multifaceted scaling of microservices with reinforcement learning,''
  \emph{IEEE Transactions on Network and Service Management}, vol.~19, no.~4,
  pp. 3995--4009, 2022.

\bibitem{qiu2020firm}
H.~Qiu, S.~S. Banerjee, S.~Jha, Z.~T. Kalbarczyk, and R.~K. Iyer,
  ``$\{$FIRM$\}$: An intelligent fine-grained resource management framework for
  $\{$SLO-Oriented$\}$ microservices,'' in \emph{14th USENIX Symposium on
  Operating Systems Design and Implementation (OSDI 20)}, 2020, pp. 805--825.

\bibitem{Zhang2020ASARSA}
S.~Zhang, T.~Wu, M.~Pan, C.~Zhang, and Y.~Yu, ``A-sarsa: A predictive container
  auto-scaling algorithm based on reinforcement learning,'' in \emph{2020 IEEE
  International Conference on Web Services (ICWS)}, 2020, pp. 489--497.

\bibitem{Rossi2019}
F.~Rossi, M.~Nardelli, and V.~Cardellini, ``Horizontal and vertical scaling of
  container-based applications using reinforcement learning,'' in \emph{2019
  IEEE 12th International Conference on Cloud Computing (CLOUD)}, 2019, pp.
  329--338.

\bibitem{XuCSUR2019}
M.~Xu and R.~Buyya, ``Brownout approach for adaptive management of resources
  and applications in cloud computing systems: A taxonomy and future
  directions,'' \emph{ACM Comput. Surv.}, vol.~52, no.~1, jan 2019.

\bibitem{TrainTicket}
X.~Zhou, X.~Peng, T.~Xie, J.~Sun, C.~Xu, C.~Ji, and W.~Zhao, ``Benchmarking
  microservice systems for software engineering research,'' in
  \emph{Proceedings of the 40th International Conference on Software
  Engineering: Companion Proceeedings, {ICSE} 2018, Gothenburg, Sweden, May 27
  - June 03, 2018}, M.~Chaudron, I.~Crnkovic, M.~Chechik, and M.~Harman,
  Eds.\hskip 1em plus 0.5em minus 0.4em\relax {ACM}, 2018, pp. 323--324.

\bibitem{locust2023}
\BIBentryALTinterwordspacing
{Locust Contributors}, ``Locust,'' 2023, accessed: 2023-06-05. [Online].
  Available: \url{https://locust.io/}
\BIBentrySTDinterwordspacing

\bibitem{guo2019limits}
J.~Guo, Z.~Chang, S.~Wang, H.~Ding, Y.~Feng, L.~Mao, and Y.~Bao, ``Who limits
  the resource efficiency of my datacenter: An analysis of alibaba datacenter
  traces,'' in \emph{Proceedings of the international symposium on quality of
  service}, 2019, pp. 1--10.

\bibitem{hochreiter1997long}
S.~Hochreiter and J.~Schmidhuber, ``Long short-term memory,'' \emph{Neural
  computation}, vol.~9, no.~8, pp. 1735--1780, 1997.

\bibitem{bucsoniu2010multi}
L.~Bu{\c{s}}oniu, R.~Babu{\v{s}}ka, and B.~De~Schutter, ``Multi-agent
  reinforcement learning: An overview,'' \emph{Innovations in multi-agent
  systems and applications-1}, pp. 183--221, 2010.

\bibitem{zimmer2014teacher}
M.~Zimmer, P.~Viappiani, and P.~Weng, ``Teacher-student framework: a
  reinforcement learning approach,'' in \emph{AAMAS Workshop Autonomous Robots
  and Multirobot Systems}, 2014.

\bibitem{hussein2017imitation}
A.~Hussein, M.~M. Gaber, E.~Elyan, and C.~Jayne, ``Imitation learning: A survey
  of learning methods,'' \emph{ACM Computing Surveys (CSUR)}, vol.~50, no.~2,
  pp. 1--35, 2017.

\bibitem{Cao2020TNSM}
Y.~{Cao}, Y.~{Zhao}, J.~{Li}, R.~{Lin}, J.~{Zhang}, and J.~{Chen},
  ``Multi-tenant provisioning for quantum key distribution networks with
  heuristics and reinforcement learning: A comparative study,'' \emph{IEEE
  Transactions on Network and Service Management}, vol.~17, no.~2, pp.
  946--957, 2020.

\bibitem{rahman2019predicting}
J.~Rahman and P.~Lama, ``Predicting the end-to-end tail latency of
  containerized microservices in the cloud,'' in \emph{2019 IEEE International
  Conference on Cloud Engineering (IC2E)}.\hskip 1em plus 0.5em minus
  0.4em\relax IEEE, 2019, pp. 200--210.

\bibitem{fujimoto2018addressing}
S.~Fujimoto, H.~Hoof, and D.~Meger, ``Addressing function approximation error
  in actor-critic methods,'' in \emph{International conference on machine
  learning}.\hskip 1em plus 0.5em minus 0.4em\relax PMLR, 2018, pp. 1587--1596.

\bibitem{GRU}
J.~Chung, C.~Gulcehre, K.~Cho, and Y.~Bengio, ``\BIBforeignlanguage{English
  (US)}{Empirical evaluation of gated recurrent neural networks on sequence
  modeling},'' in \emph{\BIBforeignlanguage{English (US)}{NIPS 2014 Workshop on
  Deep Learning, December 2014}}, 2014.

\bibitem{Xu2022TOIT}
M.~Xu, C.~Song, H.~Wu, S.~S. Gill, K.~Ye, and C.~Xu, ``Esdnn: Deep neural
  network based multivariate workload prediction in cloud computing
  environments,'' \emph{ACM Trans. Internet Technol.}, vol.~22, no.~3, aug
  2022.

\bibitem{ChainsFormer2023}
C.~Song, M.~Xu, K.~Ye, H.~Wu, S.~S. Gill, R.~Buyya, and C.~Xu, ``Chainsformer:
  A chain latency-aware resource provisioning approach for microservices
  cluster,'' in \emph{Service-Oriented Computing}, F.~Monti, S.~Rinderle-Ma,
  A.~Ruiz~Cort{\'e}s, Z.~Zheng, and M.~Mecella, Eds.\hskip 1em plus 0.5em minus
  0.4em\relax Cham: Springer Nature Switzerland, 2023, pp. 197--211.

\bibitem{carrion2022kubernetes}
C.~Carri{\'o}n, ``Kubernetes scheduling: Taxonomy, ongoing issues and
  challenges,'' \emph{ACM Computing Surveys}, vol.~55, no.~7, pp. 1--37, 2022.

\end{thebibliography}
\begin{IEEEbiography}[{\includegraphics[width=1in,height=1.25in,clip,keepaspectratio]{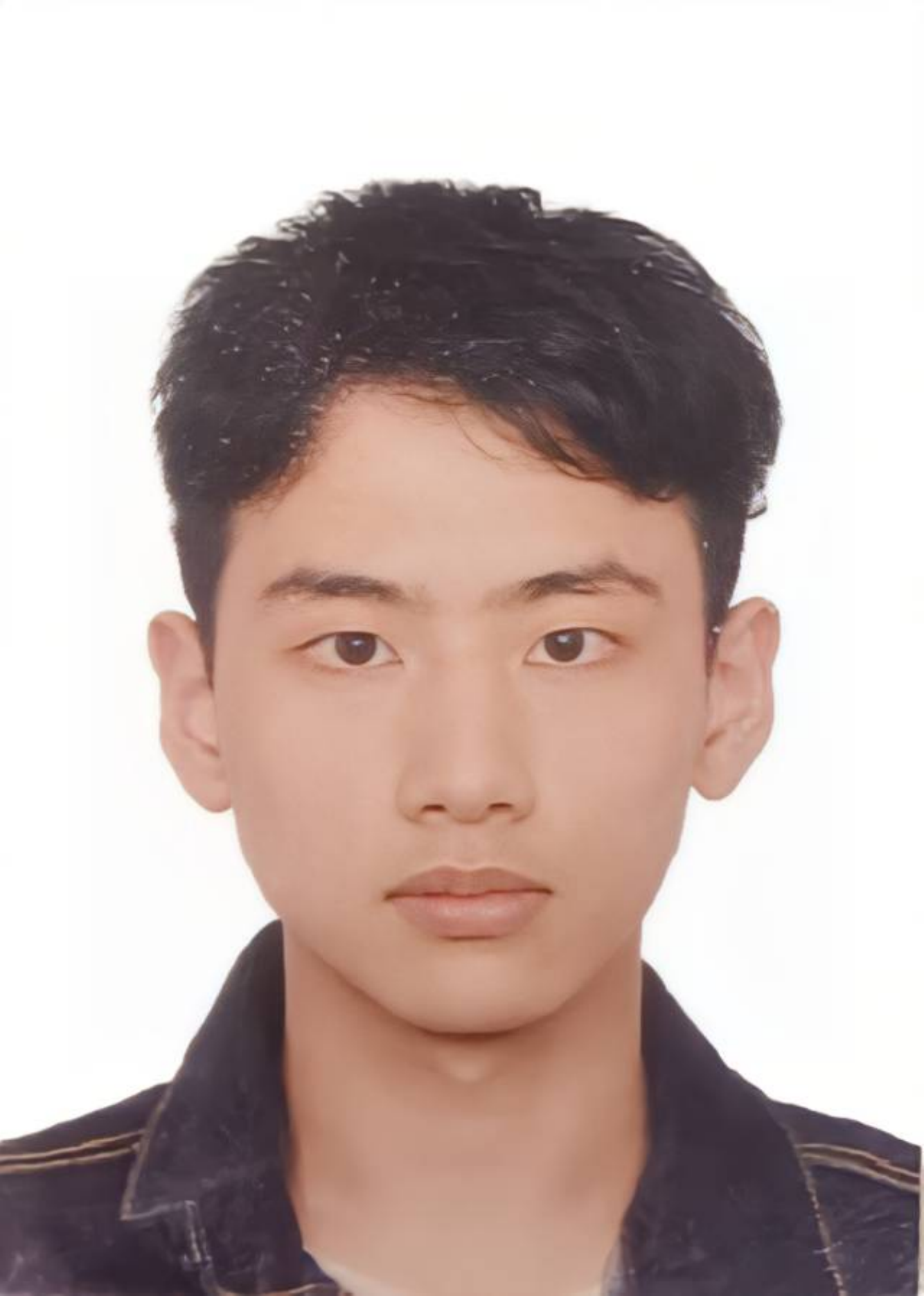}}]{Haoyu Bai}
received his BSc degree from the University of Melbourne. Now he is a PhD student at the University of Melbourne, he is also a visiting student at Shenzhen Institute of Advanced Technology, Chinese Academy of Science. His main research interest includes efficient microservice-based  application and system management.
\end{IEEEbiography}

\begin{IEEEbiography}[{\includegraphics[width=1in,height=1.25in,clip,keepaspectratio]{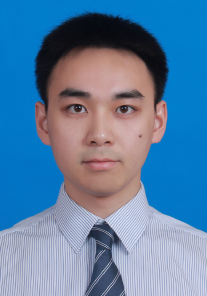}}]{Minxian Xu}
(Senior Member, IEEE) is currently an Associate Professor
at Shenzhen Institutes of Advanced Technology,
Chinese Academy of Sciences. 
his PhD degree from the University of Melbourne
in 2019. 
His research interests include resource
scheduling and optimization in cloud computing. He
has co-authored 60+ peer-reviewed papers published
in prominent international journals and conferences.
His PhD thesis was awarded the 2019 IEEE
TCSC Outstanding Ph.D. Dissertation Award. He was also awarded the 2023 IEEE TCSC Award for Excellence (Early Career Award).
\end{IEEEbiography}

\begin{IEEEbiography}[{\includegraphics[width=1in,height=1.25in,clip,keepaspectratio]{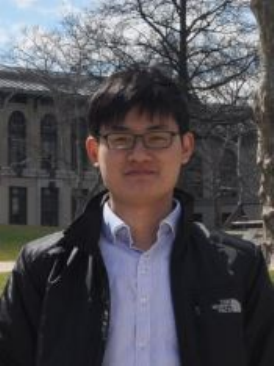}}]{Kejiang Ye}
(Senior Member, IEEE) received the BSc
and PhD degrees from Zhejiang University, in 2008
and 2013, respectively. He was also a joint PhD student
with the University of Sydney from 2012 to 2013.
After graduation, he works as post-doc researcher
with Carnegie Mellon University from 2014 to 2015
and  Wayne State University from 2015 to 2016. He is
currently a professor with the Shenzhen Institute of
Advanced Technology, Chinese Academy of Science.
His research interests focus on the performance, energy,
and reliability of cloud computing and network
systems.
\end{IEEEbiography}

\begin{IEEEbiography}[{\includegraphics[width=1in,height=1.25in,clip,keepaspectratio]{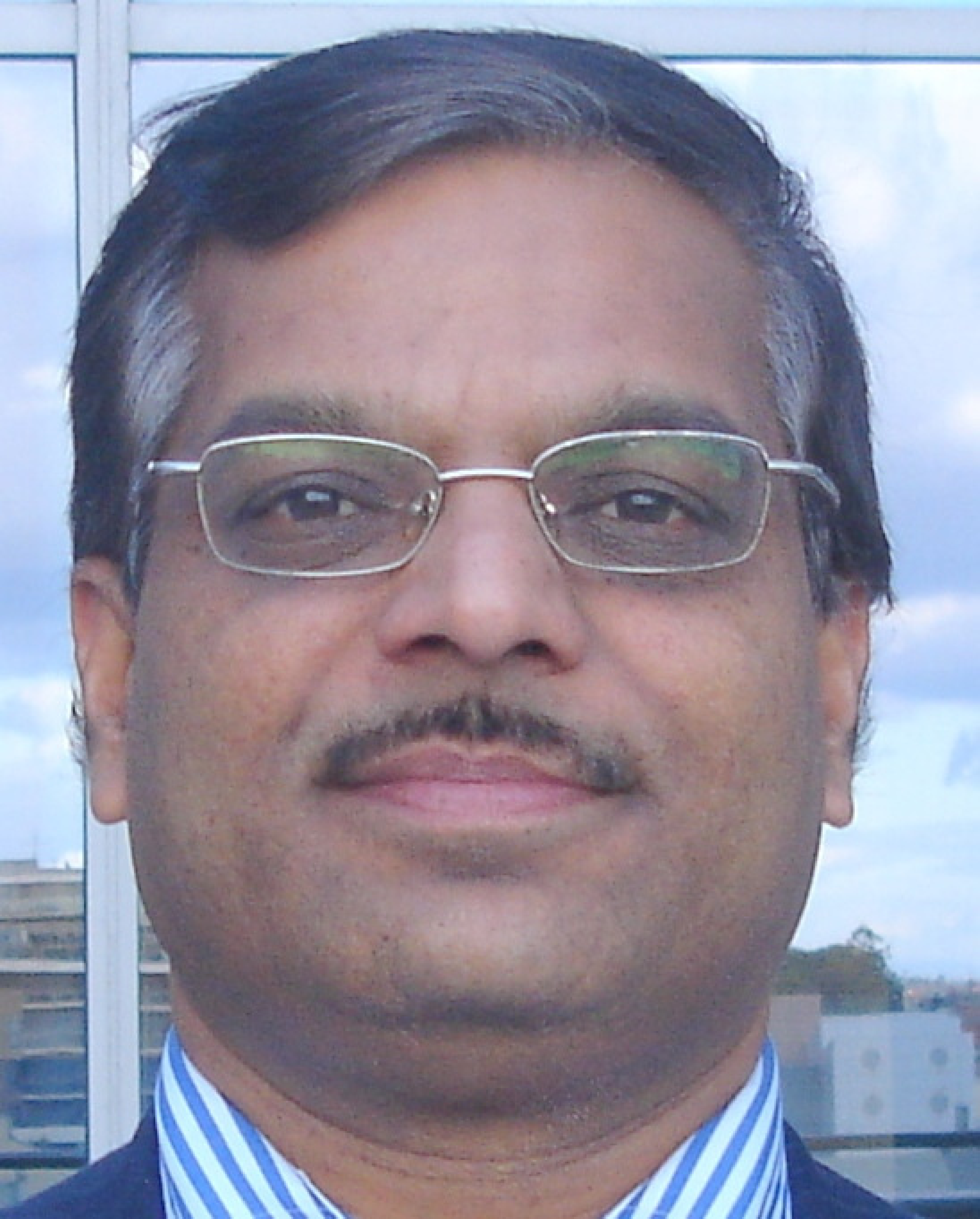}}]{Rajkumar Buyya}
(Fellow, IEEE) 
is currently a Redmond Barry distinguished professor and director with Cloud Computing and Distributed Systems (CLOUDS) Laboratory, University of Melbourne, Australia. He has authored more than 625 publications and seven textbooks including ”Mastering Cloud Computing” published by McGraw Hill, China Machine Press, and Morgan Kaufmann for Indian, Chinese and international markets, respectively. He is one of the highly cited authors in computer science and software engineering worldwide (h-index=168, g-index=322, 149,000+citations).
\end{IEEEbiography}

\begin{IEEEbiography}[{\includegraphics[width=1in,height=1.25in,clip,keepaspectratio]{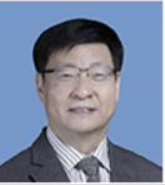}}]{Chengzhong Xu}
(Fellow, IEEE) received 
the Ph.D. degree in computer science
and engineering from the University of Hong
Kong in 1993. He is the Dean of Faculty of Science
and Technology and the Interim Director of Institute
of Collaborative Innovation, University of Macau. 
He published two research monographs
and more than 300 peer-reviewed papers in journals
and conference proceedings; his papers received about 17K citations
with an H-index of 72. His main research interests lie in parallel and distributed
computing and cloud computing. 
\end{IEEEbiography}

\end{document}